\documentclass[a4paper,11pt]{article}
\pdfoutput=1

\usepackage{jheppub}
\usepackage[T1]{fontenc}
\usepackage{hyperref}
\usepackage{graphicx}
\usepackage{amsmath}
\usepackage{epsfig}
\usepackage{subfig}
\usepackage{xcolor}
\usepackage{natbib}
\usepackage{slashed}
\usepackage{bm}

\title{Fragmentation functions for gluon into $P$-wave $B_c$ mesons}

\author{Xu-Chang Zheng and}
\author{Xing-Gang Wu}
\affiliation{Department of Physics, Chongqing Key Laboratory for Strongly Coupled Physics, \\ Chongqing University, Chongqing 401331, P.R. China}
\emailAdd{zhengxc@cqu.edu.cn, wuxg@cqu.edu.cn}

\abstract{
We calculate the fragmentation functions for a gluon into $P$-wave $B_c$ mesons within the nonrelativistic QCD factorization framework, incorporating color-singlet and color-octet contributions. Ultraviolet divergences arising from phase-space integrals are removed via operator renormalization in the modified minimal subtraction scheme. The resulting fragmentation functions are presented in both graphical form and as fitted analytic expressions.
}


\begin{document}

\maketitle

\bibliographystyle{JHEP}

\section{Introduction}

\label{intro}

The $B_c$ meson has attracted significant interest since its first observation by the CDF experiment at the Tevatron. It holds particular significance within the Standard Model (SM) as the only meson composed of two distinct heavy quark flavors. Its production mechanism differs markedly from those of charmonium and bottomonium states, as well as from hadrons containing a single heavy quark alone. Unlike the production processes for single-heavy-quark hadrons, a wealth of information about $B_c$ meson production can be predicted reliably using perturbative QCD (pQCD). The production of the $B_c$ meson in flavor-conserving collisions involves the simultaneous creation of both a $c\bar{c}$ and a $b\bar{b}$ pair, rendering its production more difficult compared to charmonium and bottomonium states. Consequently, studying the production of the $B_c$ meson offers a unique opportunity to investigate the strong interaction dynamics between quarks and gluons.

Excited $B_c$ states cannot decay via electromagnetic or strong annihilation processes, though they can undergo allowed electromagnetic or strong transitions (e.g., photon emission) to cascade down to the ground state. Consequently, excited $B_c$ states lying below the $B\bar{D}$ threshold decay (either directly or through cascades) to the ground state $B_c$ with nearly $100\%$ probability via electromagnetic or strong interactions. This decay pattern implies that when studying the inclusive production rate of the ground state $B_c$ at high-energy colliders, it is essential to account for feed-down contributions from excited states -- such as $P$-wave states. Furthermore, these $P$-wave excited $B_c$ states are directly observable and were recently detected for the first time by the LHCb collaboration~\cite{LHCb:2025uce,LHCb:2025ubr}. Studying the production of $P$-wave excited $B_c$ states is therefore not only crucial for understanding the production of the ground state $B_c$, but also valuable for probing the properties and production mechanisms of the $P$-wave states themselves.

The production cross sections of the (ground and excited) $B_c$ states can be calculated through the nonrelativistic QCD (NRQCD) factorization~\cite{nrqcd}. Within the framework of NRQCD factorization, the production cross sections are expressed as products of short-distance coefficients (SDCs) and long-distance matrix elements (LDMEs). The SDC accounts for the creation of a $(c\bar{b})$ pair with proper quantum number and can be calculated perturbatively as a series in the strong coupling constant. The LDME, on the other hand, characterizes the nonperturbative transition of the $(c\bar{b})$ pair into a physical $B_c$ state, and can be obtained via the potential model calculation or extracted from experimental data. NRQCD factorization has been widely applied to predict the production rates of the $B_c$ states. Moreover, two generators, BCVEGPY~\cite{Chang:2015qea, Chang:2006xka, Chang:2005hq, Chang:2003cq} and BEEC~\cite{Yang:2022zpc,Yang:2013vba} (both based on the NRQCD factorization), have been developed to simulate the production of the $B_c$ states at hadron and electron-positron colliders, respectively. However, when the transverse momentum ($p_T$) of the $B_c$ state is very large compared to the mass ($m_{(c\bar{b})}$) of the $B_c$ state, large logarithms of the form ${\rm ln}(p_T^2/m_{(c\bar{b})}^2)$, which arise from the emission of collinear gluons or light-quark pairs, will appear in the perturbative series of the SDCs. These large logarithms may spoil the convergence of the fixed-order perturbative series. 

At the large $p_T$ region, the production cross sections of the $B_c$ states can also be calculated through the fragmentation function approach. Under this approach, the production cross section of the $B_c$ state ($H_{c\bar{b}}$) can be written as~\cite{Collins:1989gx}: 
\begin{eqnarray}
d\sigma_{A+B \to H_{c\bar{b}}(p_T)+X}=&& \sum_i d \hat{\sigma}_{A+B\to i+X}(p_T/z,\mu_F) \otimes D_{i\to H_{c\bar{b}}}(z,\mu_F)+{\cal O}(m_{(c\bar{b})}^2/p_T^2),
\label{pqcd-fact}
\end{eqnarray}
where $d \hat{\sigma}_{A+B\to i+X}(p_T/z,\mu_F)$ is the inclusive cross section for producing a parton $i$, and $D_{i\to H_{c\bar{b}}}(z,\mu_F)$ is the fragmentation function of the parton $i$ into $H_{c\bar{b}}$. $\mu_F$ is the factorization scale, and $\otimes$ stands for a convolution integral over $z$. Within the fragmentation function approach, the large logarithms arising from the collinear emissions can be resummed through solving the evolution equation of the fragmentation function.

The fragmentation function $D_{i\to H_{c\bar{b}}}(z,\mu_F)$ is a central ingredient of the fragmentation function formalism. In contrast to the case of light hadrons, whose fragmentation functions are intrinsically nonperturbative, those associated with $B_c$ mesons contain short-distance perturbative components and can be systematically evaluated within the NRQCD factorization framework. Accordingly, they can be expressed as
\begin{eqnarray}
D_{i\to H_{c\bar{b}}}(z,\mu_F)=\sum_n d_{i\to (c\bar{b})[n]}(z,\mu_F) \langle {\cal O}^{H_{c\bar{b}}}(n) \rangle,
\label{frag-nrqcd}
\end{eqnarray}
where $d_{i\to (c\bar{b})[n]}(z,\mu_F)$ denotes the SDC, and $\langle {\cal O}^{H_{c\bar{b}}}(n) \rangle$ represents the corresponding LDME. 

The leading-order (LO) fragmentation functions for a heavy quark ($c$ or $\bar{b}$) into $S$-wave~\cite{Chang:1992bb, Braaten:1993jn, Ma:1994zt}, $P$-wave~\cite{Chen:1993ii, Yuan:1994hn}, and $D$-wave~\cite{Cheung:1995ir} $B_c$ states were first calculated in the 1990s. Recently, the fragmentation functions for a heavy quark fragmenting into $S$-wave $B_c$ states have been computed up to next-to-leading order (NLO) in both $\alpha_s$~\cite{Zheng:2019gnb} and $v$~\cite{Yang:2019gga}, where $v$ denotes the relative velocity of the heavy quarks in the rest frame of the meson. Furthermore, the fragmentation functions, which start at order $\alpha_s^3$, for a gluon fragmenting into $S$-wave $B_c$ states have also been calculated in refs.~\cite{Zheng:2021sdo, Feng:2021qjm}. These $S$-wave $B_c$ fragmentation functions up to order $\alpha_s^3$ have been applied in studies of $B_c$ production processes~\cite{Celiberto:2024omj, Zheng:2023atb, Dadfar:2022nct, Celiberto:2022keu, Zhang:2021ypo, Zheng:2019egj}. To date, fragmentation functions for $P$-wave $B_c$ states are only available at order $\alpha_s^2$. As emphasized earlier, studying the production of $P$-wave $B_c$ mesons is important for understanding the production mechanisms of both ground and excited $B_c$ states. In this paper, as a follow-up to our previous work of ref.~\cite{Zheng:2021sdo}, we calculate the fragmentation functions for a gluon fragmenting into $P$-wave $B_c$ states, which also starts at order $\alpha_s^3$.

The remainder of this paper is organized as follows. Section~\ref{calcmethod} introduces the gauge-invariant definition of the gluon fragmentation function and the NRQCD factorization formalism for the fragmentation functions of the $P$-wave $B_c$ production. Numerical results are presented in section~\ref{numer}, and section~\ref{sum} contains a summary.

\section{Calculation method}
\label{calcmethod}

\subsection{Definition of the gluon fragmentation function}

Before proceeding with the calculation, we present the definition of the fragmentation function for a gluon fragmenting into a generic hadron $H$ for the sake of self-consistency. 

It is convenient to use the light-cone coordinates in the definition and calculation of the fragmentation function. For a $d$-dimensional vector $V^{\mu}$, its light-cone coordinates are defined as $V^{\mu}=(V^+,V^-,\textbf{V}_{\perp})= \left((V^0+V^{d-1})/\sqrt{2}, (V^0-V^{d-1})/\sqrt{2}, \textbf{V}_{\perp}\right)$, and the scalar product of two vectors $V$ and $W$ becomes $V \cdot W = V^+ W^- +V^- W^+ -\textbf{V}_{\perp}\cdot \textbf{W}_{\perp}$. Then the fragmentation function for a gluon fragmenting into a hadron $H$ is defined as
\begin{eqnarray}
D_{g\to H}(z)=&&\frac{-g_{\mu \nu}\,z^{d-3}}{2\pi K^+ (N_c^2-1)(d-2)}\sum_{X} \int_{-\infty}^{+\infty} dx^- e^{-iK^+ x^-} \langle 0 \vert G_c^{+\mu}(0) \mathcal{E}^{\dagger}(0,0,\textbf{0}_{\perp})_{cb}\nonumber \\
&&\times  \vert H(P^+,{\bf 0}_\perp)+X \rangle \langle H(P^+,{\bf 0}_\perp)+X\vert \mathcal{E}(0,x^-,\textbf{0}_{\perp})_{ba} G_a^{+\nu}(0,x^-,\textbf{0}_{\perp}) \vert 0\rangle.
\label{defrag1}
\end{eqnarray}
Here we have adopted the gauge-invariant operator definition of the fragmentation function, first proposed by Collins and Soper~\cite{Collins:1981uw}. $G_c^{\mu \nu}$ is the gluon field-strength operator, $K$ and $P$ are momenta of the initial gluon and the produced hadron $H$, respectively. $z\equiv P^+/K^+$ is the  fraction of the gluon’s plus-momentum carried by the produced hadron. The gauge link $\mathcal{E}(0,x^-,\textbf{0}_{\perp})_{ba}$ is
\begin{eqnarray}
\mathcal{E}(0,x^-,\textbf{0}_{\perp})_{ba}={\cal P}{\rm exp}\left[ ig_s \int_{x^-}^{\infty}dz^- A^+(0,z^-,\textbf{0}_{\perp})  \right]_{ba},
\end{eqnarray}
where ${\cal P}$ indicates path ordering, $[A^{\mu}(x)]_{ba}=-if^{cba}A^{\mu}_c(x)$ is the matrix-valued gluon
field in the adjoint representation. 

The above definition (eq.~(\ref{defrag1})) is formulated in a reference frame with zero transverse momentum for the produced hadron $H$. In this reference frame, it is convenient to introduce a lightlike vector $n^{\mu}$ defined as $n^{\mu}=(0,1,\textbf{0}_{\perp})$. The plus component of an arbitrary momentum $p$ can then be written in a manifestly Lorentz-invariant form as $p^+=p\cdot n$. Following this definition, the Feynman rules for calculating the gluon fragmentation function can be derived straightforwardly. The standard QCD Feynman rules also apply, while additional rules specific to the gluon fragmentation process are summarized below. 

The fragmentation function can be expressed as a sum over cut diagrams, each containing an eikonal line (from the gauge link) that connects the operators on either side of the cut. An overall factor,
\begin{eqnarray}
N_{\rm CS}=\frac{1}{(N_c^2-1)(d-2)}\frac{z^{d-3}}{2\pi K\cdot n},
\label{eq.Ncs}
\end{eqnarray}
arises from the definition in $d$-dimensional spacetime. In the following, we provide the Feynman rules only for the elements on the left side of the cut; those on the right side can be obtained by taking the complex conjugate. The operator vertex that creates a gluon line along with an eikonal line contributes a factor
\begin{eqnarray}
-i(K \cdot n \, g^{\mu \lambda} - Q^{\mu} n^{\lambda}) \delta_{ac},
\end{eqnarray}
where $Q$ is the momentum of the created gluon, $K$ is the sum of the momenta of the gluon line and the eikonal line, $\lambda$ is the gluon's Lorentz index, and $a, c$ are the respective color indices of the eikonal line and the gluon. The eikonal-line-gluon interaction vertex contributes a factor $g_s f^{abc} n^{\mu}$, with $\mu$ and $a$ representing the Lorentz and color indices of the gluon, and $b, c$ representing the color indices of the eikonal line on the left and right sides, respectively. The propagator for an eikonal line with momentum $l$ is given by $i \delta_{ab}/(l \cdot n + i \epsilon)$. And when an eikonal line with momentum $l$ is cut, it will introduce a factor of $2 \pi \delta(l \cdot n)$.

\subsection{NRQCD factorization for the fragmentation functions}

At the LO in $v$, the NRQCD factorization formalism for the fragmentation functions of a gluon into $P$-wave $B_c$ states can be expressed as
\begin{eqnarray}
D_{g \to B_c(n^1P_1)}(z,\mu_F)=&& d_{g\to (c\bar{b})[^1P^{[1]}_1]}(z,\mu_F) \langle {\cal O}^{B_c(n^1P_1)}(^1P^{[1]}_1) \rangle\nonumber \\
                             && +d_{g\to (c\bar{b})[^1S^{[8]}_0]}(z,\mu_F) \langle {\cal O}^{B_c(n^1P_1)}(^1S^{[8]}_0) \rangle, \label{nrqcd-fact-1} \\ 
D_{g \to B_c(n^3P_J)}(z,\mu_F)=&& d_{g\to (c\bar{b})[^3P^{[1]}_J]}(z,\mu_F) \langle {\cal O}^{B_c(n^3P_J)}(^3P^{[1]}_J) \rangle\nonumber \\
                             && +d_{g\to (c\bar{b})[^3S^{[8]}_1]}(z,\mu_F) \langle {\cal O}^{B_c(n^3P_J)}(^3S^{[8]}_1) \rangle,\label{nrqcd-fact-2}
\end{eqnarray}
where $n=2,3,\cdots,$ denotes the principal quantum number of the $P$-wave states\footnote{We adopt $B_c(n^{2S+1}P_J)$ to denote the $P$-wave $B_c$ state whose dominant Fock state is $(c\bar{b})[n^{2S+1}P_J]$.}, $J=0,1,2$ denotes the total angular momentum quantum number of the spin-triplet $P$-wave states. There are two distinct mechanisms that contribute to the $P$-wave $B_c$ fragmentation functions at the LO in $v$, they are referred to as the color-singlet mechanism and the color-octet mechanism. As shown in eqs. (\ref{nrqcd-fact-1}) and (\ref{nrqcd-fact-2}), the SDCs do not depend on the principal quantum number $n$. This independence naturally arises from the scale separation inherent in the NRQCD factorization framework. The SDCs describe the perturbative creation of the $(c\bar{b})$ pair which occurs at a short distance scale of roughly $1/m_{b,c}$. At this energy scale, the heavy quark pair is essentially a free perturbative state and remains insensitive to the nonperturbative binding dynamics. The binding process that determines the specific radial excitation state (characterized by $n$) occurs at a much larger distance scale of $1/(m_{b,c} v)$. Consequently, all physical dependencies on the principal quantum number $n$ are completely absorbed into the LDMEs. 

The LDMEs are defined though the local four-fermion operators~\cite{nrqcd}, i.e.,
\begin{eqnarray}
\langle {\cal O}^{B_c(n^1P_1)}(^1P^{[1]}_1) \rangle=&&\langle 0 \vert \chi_b^{\dagger}(-\frac{i}{2}\overleftrightarrow{D^i})\psi_c \sum_{X} \vert B_c(n^1P_1)+X  \rangle \nonumber \\
&& \cdot  \langle B_c(n^1P_1)+X  \vert \psi_c^{\dagger}(-\frac{i}{2}\overleftrightarrow{D^i})\chi_b \vert 0 \rangle,\\
\langle {\cal O}^{B_c(n^3P_0)}(^3P^{[1]}_0) \rangle=&&\frac{1}{3}\langle 0 \vert \chi_b^{\dagger}(-\frac{i}{2}\overleftrightarrow{\bm{D}}\cdot \bm{\sigma})\psi_c \sum_{X} \vert B_c(n^3P_0)+X  \rangle \nonumber \\
&& \cdot \langle B_c(n^3P_0)+X  \vert \psi_c^{\dagger}(-\frac{i}{2}\overleftrightarrow{\bm{D}}\cdot \bm{\sigma})\chi_b \vert 0 \rangle,\\
\langle {\cal O}^{B_c(n^3P_1)}(^3P^{[1]}_1) \rangle=&&\frac{1}{2}\langle 0 \vert \chi_b^{\dagger}(-\frac{i}{2}\overleftrightarrow{\bm{D}}\times \bm{\sigma})^i \psi_c \sum_{X} \vert B_c(n^3P_1)+X  \rangle \nonumber \\
&& \cdot \langle B_c(n^3P_1)+X  \vert \psi_c^{\dagger}(-\frac{i}{2}\overleftrightarrow{\bm{D}}\times \bm{\sigma})^i \chi_b \vert 0 \rangle,\\
\langle {\cal O}^{B_c(n^3P_2)}(^3P^{[1]}_2) \rangle=&&\langle 0 \vert \chi_b^{\dagger}(-\frac{i}{2}\overleftrightarrow{D}^{(i} \sigma^{j)}) \psi_c \sum_{X} \vert B_c(n^3P_2)+X  \rangle \nonumber \\
&& \cdot \langle B_c(n^3P_2)+X  \vert \psi_c^{\dagger}(-\frac{i}{2}\overleftrightarrow{D}^{(i} \sigma^{j)}) \chi_b \vert 0 \rangle,\\
\langle {\cal O}^{B_c(n^1P_1)}(^1S^{[8]}_0) \rangle=&&\langle 0 \vert \chi_b^{\dagger}T^a \psi_c \sum_{X} \vert B_c(n^1P_1)+X  \rangle \nonumber \\
&& \cdot  \langle B_c(n^1P_1)+X  \vert \psi_c^{\dagger}T^a\chi_b \vert 0 \rangle,\\
\langle {\cal O}^{B_c(n^3P_J)}(^3S^{[8]}_1) \rangle=&&\langle 0 \vert \chi_b^{\dagger}\sigma^i T^a \psi_c \sum_{X} \vert B_c(n^3P_J)+X  \rangle \nonumber \\
&& \cdot  \langle B_c(n^3P_J)+X  \vert \psi_c^{\dagger} \sigma^i T^a\chi_b \vert 0 \rangle,
\end{eqnarray}
where $\psi_c$ is the Pauli spinor field that annihilates a charm quark, and $\chi_b$ is the Pauli spinor field that creates a bottom antiquark. The symmetric traceless component of a tensor is defined as $A^{(ij)}=(A^{ij}+A^{ji})/2-A^{kk}\delta^{ij}/3$. 

According to the heavy quark spin symmetry (HQSS)~\cite{nrqcd}, the LDMEs have the following relations:
\begin{eqnarray}
\langle {\cal O}^{B_c(n^3P_J)}(^3P^{[1]}_J) \rangle &=& \frac{(2J+1)}{3}\langle {\cal O}^{B_c(n^1P_1)}(^1P^{[1]}_1) \rangle ,\nonumber \\
\langle {\cal O}^{B_c(n^3P_J)}(^3S^{[8]}_1) \rangle &=& \frac{(2J+1)}{3} \langle {\cal O}^{B_c(n^1P_1)}(^1S^{[8]}_0) \rangle.
\label{eq.HQSS}
\end{eqnarray}
These relations are valid up to corrections of relative order $v^2$. Through the application of these approximate HQSS relations, the number of input nonperturbative LDMEs in eqs.~(\ref{nrqcd-fact-1}) and (\ref{nrqcd-fact-2}) is reduced to two.

To order $\alpha_s$, the color-singlet LDME $\langle {\cal O}^{B_c(n^1P_1)}(^1P^{[1]}_1) \rangle$ is scale invariant and it can be related to the derivative of the radial wave function at the origin for the $P$-wave $B_c$ states:
\begin{eqnarray}
\langle {\cal O}^{B_c(n^1P_1)}(^1P^{[1]}_1) \rangle \approx \frac{9N_c}{2\pi} \vert R'_{nP}(0) \vert^2.
\label{eq.csLDMEs}
\end{eqnarray}
The color-octet LDME $\langle {\cal O}^{B_c(n^1P_1)}(^1S^{[8]}_0) \rangle$ satisfies the renormalization-group equation (RGE)~\cite{Yuan:1994hn}:
\begin{eqnarray}
\mu_{\Lambda}\frac{d}{d \mu_{\Lambda}}\langle {\cal O}^{B_c(n^1P_1)}(^1S^{[8]}_0) \rangle=\frac{4\alpha_s(\mu_{\Lambda})}{27\pi m_{\rm red}^2}\langle {\cal O}^{B_c(n^1P_1)}(^1P^{[1]}_1)\rangle,
\end{eqnarray}
where $\mu_{\Lambda}$ is the factorization scale in the NRQCD factorization, and the reduced mass $m_{\rm red}=m_b\, m_c/(m_b+m_c)$. To order $\alpha_s$, the solution to this RGE is
\begin{eqnarray}
\langle {\cal O}^{B_c(n^1P_1)}(^1S^{[8]}_0) \rangle_{\mu_{\Lambda}}=\langle {\cal O}^{B_c(n^1P_1)}(^1S^{[8]}_0) \rangle_{\mu_{\Lambda 0}}+\frac{8}{27\beta_0 m_{\rm red}^2}{\rm ln}\left( \frac{\alpha_s(\mu_{\Lambda 0})}{\alpha_s(\mu_{\Lambda})}\right)\langle {\cal O}^{B_c(n^1P_1)}(^1P^{[1]}_1)\rangle, \nonumber \\
\label{eq.evolsol}
\end{eqnarray}
where $\beta_0=11-2n_f/3$ is the first coefficient of the QCD $\beta$ function, $n_f$ denotes the number of the active quark flavors, and $\mu_{\Lambda 0}$ is an initial NRQCD factorization scale.

\subsection{Calculation of SDCs of the fragmentation functions}

In order to obtain the SDCs for these fragmentation functions, we first apply the NRQCD factorization formalism to the fragmentation functions for producing an on-shell $(c\bar{b})$ pair with proper quantum number, i.e.,
\begin{eqnarray}
D_{g \to (c\bar{b})[n]}(z,\mu_F)= \sum_{n'} d_{g\to (c\bar{b})[n']}(z,\mu_F) \langle {\cal O}^{(c\bar{b})[n]}(n') \rangle. 
\label{nrqcd-fact-3} 
\end{eqnarray}
where the fragmentation functions for an on-shell $(c\bar{b})$ pair on the left hand side of eq.~(\ref{nrqcd-fact-3}) are computed through perturbative QCD, and the LDMEs on the right hand side of eq.~(\ref{nrqcd-fact-3}) are computed through perturbative NRQCD. Then, the SDCs of the fragmentation functions can be extracted by comparing both sides of eq.~(\ref{nrqcd-fact-3}). Since the SDCs are insensitive to the long-distance dynamics, those determined through the fragmentation functions for an on-shell $(c\bar{b})$ pair can be directly applied to calculate the fragmentation functions for physical $P$-wave $B_c$ states. In the present work, we only calculated the fragmentation functions for gluon into $P$-wave $B_c$ states at the LO in $\alpha_s$. At the LO level, only the term of $n'=n$ on the right side of eq.~(\ref{nrqcd-fact-3}) is non-vanishing. Therefore, the involved $n$ in our calculation are $^1P^{[1]}_1$, $^3P^{[1]}_J(J=0,1,2)$, $^1S^{[8]}_0$ and $^3S^{[8]}_1$, respectively.

\begin{figure}[htbp]
\centering
\includegraphics[width=0.6\textwidth]{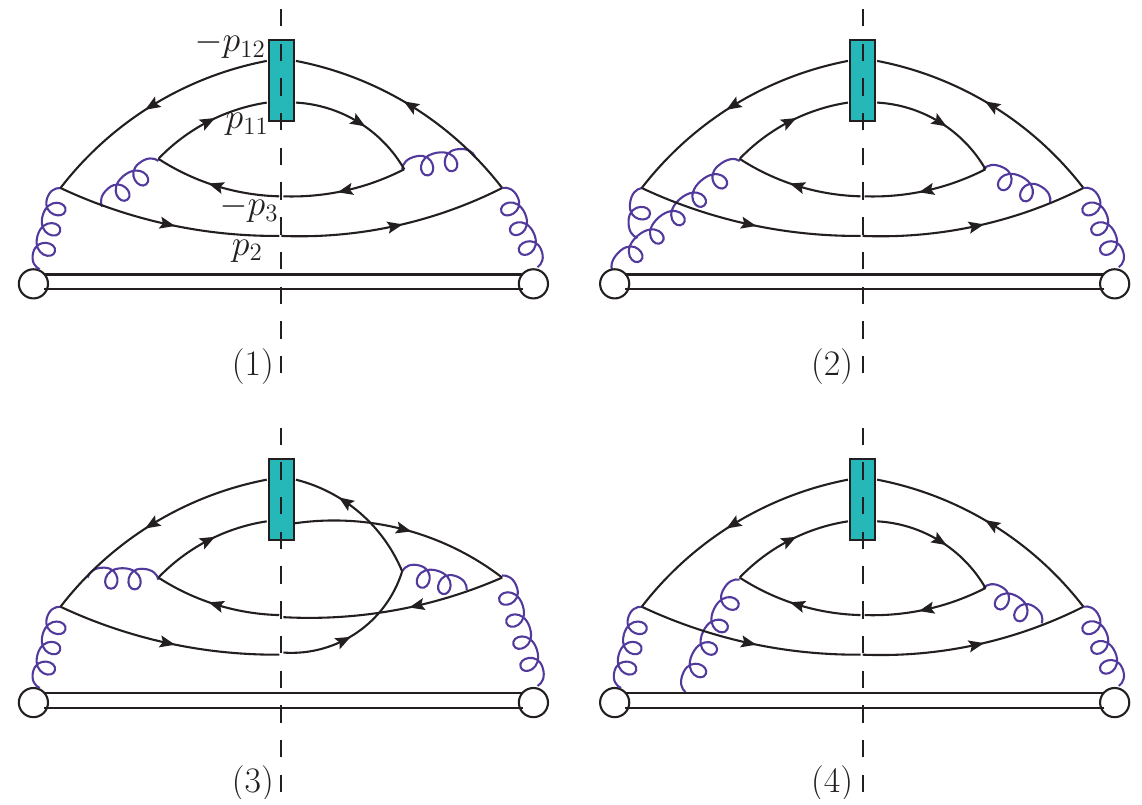}
\caption{Four sample cut diagrams for the fragmentation functions $D_{g \to (c\bar{b})[n]}(z,\mu_F)$.
 } \label{feyn}
\end{figure}

There are $49$ cut diagrams that contribute to the fragmentation functions of a gluon into $P$-wave $(c\bar{b})$ pair, e.g. $D_{g \to (c\bar{b})[n]}(z,\mu_F)$, in the Feynman gauge, with four sample diagrams given in Fig.~\ref{feyn}. Using the Feynman rules, the squared amplitude for the fragmentation function can be written down directly. For the calculation, we utilize the FeynCalc package~\cite{Shtabovenko:2023idz,Mertig:1990an} to handle the Dirac and color algebra. And in calculating the fragmentation functions $D_{g \to (c\bar{b})[n]}(z,\mu_F)$, it is convenient to use the projection operators for the spin and color states of the $(c\bar{b})$ pair \cite{Petrelli:1997ge,Bodwin:2010fi}. For the spin singlet, the projection operator reads
\begin{eqnarray}
\Pi_1= \frac{\sqrt{M}}{{4{m_b}{m_c}}}(\slashed{p}_{12}- m_b) \gamma_5 (\slashed{p}_{11} + m_c).
\label{eq.Pi1}
\end{eqnarray}
For the spin triplet, the projection operator takes the form
\begin{eqnarray}
\Pi_3^{\alpha}= \frac{\sqrt{M}}{{4{m_b}{m_c}}}(\slashed{p}_{12}- m_b) \gamma^{\alpha} (\slashed{p}_{11} + m_c),
\label{eq.Pi3}
\end{eqnarray}
where $M=m_b+m_c$ and $r_c=(1-r_b)=m_c/M$. The momenta are given by $p_{11}=r_c\,p_1+q$, $p_{12}=r_b\,p_1-q$, where $p_1$ represents the momentum of the produced $(c\bar{b})$ pair. 
The color projection operators for the color singlet and octet are
\begin{eqnarray}
\Lambda_1 &=& \frac{1}{\sqrt{3}} \textbf{1},\nonumber \\
\Lambda^a_8 &=& \sqrt{2}\, T^a,
\end{eqnarray}
where $\textbf{1}$ and $T^a$ denote the unit matrix and a generator of the fundamental representation of the $SU(3)$ group, respectively.

The squared amplitudes take the form:
\begin{eqnarray}
{\cal A}(^1S_0^{[8]}) &=& {\rm Tr}\left( \Pi_1 \Lambda_8^a \, \Gamma \, \bar{\Pi}_1 \Lambda_8^a \, \Gamma' \right) \vert_{q=\bar{q}=0}, \\
{\cal A}(^3S_1^{[8]}) &=& \sum_{J_z}\epsilon^*_{\alpha}\epsilon^{ }_{\beta} {\rm Tr}\left( \Pi_3^{\alpha} \Lambda_8^a \, \Gamma \, \bar{\Pi}_3^{\beta} \Lambda_8^a \, \Gamma' \right) \vert_{q=\bar{q}=0}, \\
{\cal A}(^1P_1^{[1]}) &=& \sum_{J_z} \epsilon^*_{\alpha}\epsilon^{ }_{\beta} {\bm q}^2\left[ \frac{\partial^2}{\partial \bar{q}_{\beta} \partial q_{\alpha}} {\rm Tr}\left( \Pi_1 \Lambda_1 \, \Gamma \, \bar{\Pi}_1 \Lambda_1 \, \Gamma' \right) \big\vert_{q=\bar{q}=0}\right], \\
{\cal A}(^3P_J^{[1]}) &=& \sum_{J_z} \epsilon^{(J)*}_{\alpha \beta}\epsilon^{(J) }_{\alpha' \beta'} {\bm q}^2\left[ \frac{\partial^2}{\partial \bar{q}_{\beta'} \partial q_{\beta}} {\rm Tr}\left( \Pi_3^{\alpha} \Lambda_1 \, \Gamma \, \bar{\Pi}_3^{\alpha'} \Lambda_1 \, \Gamma' \right) \big\vert_{q=\bar{q}=0}\right],
\end{eqnarray}
where $\Gamma$ and $\Gamma'$ both stand for strings of Dirac $\gamma$ and color matrices, ${\bm q}$ denotes the 3-momentum component of the relative 4-momentum $q$ in the rest frame of the $(c\bar{b})$ pair, $J_z$ denotes the projection of the total angular momentum on an axis, and $\epsilon^*_{\alpha}$ and $\epsilon^{(J)*}_{\alpha \beta}$ denote the polarization vector and tensor of the produced $(c\bar{b})$ pair, respectively. It is worth noting that, following the standard convention in the literature of fragmentation function calculations (e.g., refs.~\cite{Bodwin:2012xc,Artoisenet:2014lpa,Artoisenet:2018dbs}), we use the symbol $\mathcal{A}$ to denote the squared amplitude (i.e., the evaluated trace of the cut diagrams) rather than the amplitude itself. In order to simultaneously calculate the first derivative of amplitude and complex conjugate amplitude with respect to relative momentum, we have replaced the relative momentum $q$ in the complex conjugate amplitude with $\bar{q}$. The conjugate projection operators $\bar{\Pi}_1$ and $\bar{\Pi}_3^{\beta}$ are 
\begin{eqnarray}
\bar{\Pi}_1 &=& \gamma^0 \Pi_1^{\dagger} \gamma^0= -\frac{\sqrt{M}}{{4{m_b}{m_c}}} (\slashed{p}_{11} + m_c) \gamma_5 (\slashed{p}_{12}- m_b), \\
\bar{\Pi}_3^{\alpha} &=& \gamma^0 \Pi_3^{\alpha \dagger} \gamma^0 =\frac{\sqrt{M}}{{4{m_b}{m_c}}}(\slashed{p}_{11} + m_c) \gamma^{\alpha} (\slashed{p}_{12}- m_b).
\end{eqnarray}

The sum over polarizations is given by:
\begin{eqnarray}
\sum_{J_z} \epsilon^*_{\alpha}\epsilon^{ }_{\beta} &=& \Pi_{\alpha\beta},\\
\epsilon^{(0)*}_{\alpha \beta}\epsilon^{(0) }_{\alpha' \beta'} &=& \frac{1}{d-1}\Pi_{\alpha \beta}\Pi_{\alpha' \beta'},\\
\sum_{J_z} \epsilon^{(1)*}_{\alpha \beta}\epsilon^{(1) }_{\alpha' \beta'} &=& \frac{1}{2}\left[  \Pi_{\alpha \alpha'}\Pi_{ \beta\beta'}- \Pi_{\alpha \beta'}\Pi_{\alpha' \beta}  \right],\\
\sum_{J_z} \epsilon^{(2)*}_{\alpha \beta}\epsilon^{(2) }_{\alpha' \beta'} &=& \frac{1}{2}\left[  \Pi_{\alpha \alpha'}\Pi_{ \beta\beta'}+ \Pi_{\alpha \beta'}\Pi_{\alpha' \beta}  \right]- \frac{1}{d-1}\Pi_{\alpha \beta}\Pi_{\alpha' \beta'},
\end{eqnarray}
where
\begin{eqnarray}
 \Pi_{\alpha\beta}=-g_{\alpha\beta}+\frac{p_{1\alpha}p_{1\beta}}{p_1^2}.
\end{eqnarray}

Once the squared amplitude is obtained, the contributions of the $49$ cut diagrams to the fragmentation functions can be expressed as
\begin{eqnarray}
D^{(1-49)}_{g \to (c\bar{b})[n]}(z)
= N_{\rm CS}\int d\phi_{3}(p_1,p_2,p_3) {\cal A}_{(1-49)},
\label{cut-contribution}
\end{eqnarray}
where $N_{\rm CS}$ represents the overall factor defined in eq.~(\ref{eq.Ncs}), ${\cal A}_{(1-49)}$ denotes the squared amplitude summed over all 49 cut diagrams, and $d\phi_{3}(p_1,p_2,p_3)$ is the three-body phase-space element, given explicitly by
\begin{eqnarray}
d\phi_{3}(p_1,p_2,p_3)
= 2\pi \, \delta\left(K^+ - \sum_{i=1}^{3} p_i^+\right) \mu^{2(4-d)} \prod_{i=2,3}\frac{\theta(p_i^+) dp_i^+}{4\pi p_i^+} \frac{d^{d-2}\mathbf{p}_{i\perp}}{(2\pi)^{d-2}}.
\end{eqnarray}

When evaluated in four dimensions, the integral in eq.~(\ref{cut-contribution}) exhibits ultraviolet (UV) divergences. These divergences originate from the regions of phase space where the transverse momenta satisfy $|\mathbf{p}_{2\perp}|\to\infty$ or $|\mathbf{p}_{3\perp}|\to\infty$. To regulate such divergences, we employ dimensional regularization by working in $d=4-2\epsilon$ dimensions, in which the UV divergences manifest themselves as poles in the regulator $\epsilon$.

A direct evaluation of the integral in eq.~(\ref{cut-contribution}) in $d$ dimensions is impractical due to the complicated structure of ${\cal A}_{(1-49)}$. To overcome this difficulty, we adopt the subtraction method, which has been successfully applied in recent calculations of quarkonium fragmentation functions~\cite{Zheng:2019gnb,Zheng:2021sdo,Zheng:2021ylc,Zheng:2019dfk,Artoisenet:2014lpa}. Within this framework, the contribution from the 49 cut diagrams is rewritten as
\begin{eqnarray}
D^{(1-49)}_{g \to (c\bar{b})[n]}(z)
&=& N_{\rm CS}\int d\phi_{3}(p_1,p_2,p_3)
\left[{\cal A}_{(1-49)}-{\cal A}_{S}\right] \nonumber \\
&& + N_{\rm CS}\int d\phi_{3}(p_1,p_2,p_3){\cal A}_{S},
\label{sub-method}
\end{eqnarray}
where ${\cal A}_{S}$ is an appropriately constructed subtraction term that reproduces the same singular behavior as ${\cal A}_{(1-49)}$. The first integral in eq.~(\ref{sub-method}) is free of divergences and can therefore be evaluated numerically in four dimensions. The second integral contains all the UV singularities of the original expression in eq.~(\ref{cut-contribution}) and must be computed analytically in $d$ dimensions. More details on the construction and integration of subtraction terms can be found in ref.~\cite{Zheng:2021sdo}.

To eliminate the UV divergences in  $D^{(1-49)}_{g \to (c\bar{b})[n]}(z)$, the renormalization of the operators~\cite{Mueller:1978xu} defining the fragmentation functions are performed. This renormalization is carried out within the modified minimal subtraction scheme ($\overline{\rm MS}$). As a result, the fragmentation functions are derived as follows:
\begin{eqnarray}
D_{g \to (c\bar{b})[n]}(z,\mu_F)
&=& D^{(1-49)}_{g \to (c\bar{b})[n]}(z)-\frac{\alpha_s}{2\pi}\left[\frac{1}{\epsilon_{_{UV}}}- \gamma_E + {\rm ln}\frac{4\pi \mu^2}{\mu_F^2} \right]\nonumber \\
&& ~~~ \times \int_z^1 \frac{dy}{y}\Bigg[\sum_{Q=\bar{b},c}P_{Qg}(y) D_{Q\to (c\bar{b})[n]}^{\rm LO}(z/y)\Bigg],
\label{FFRen}
\end{eqnarray}
where $D_{\bar{b}\to (c\bar{b})[n]}^{\rm LO}(z)$ and $D_{c\to (c\bar{b})[n]}^{\rm LO}(z)$ represent the LO fragmentation functions in $d$ dimensions. The splitting function for $g\to Q$ is given by
\begin{equation}
P_{Qg}(y)=T_F\left[y^2+(1-y)^2\right],
\label{lospfun}
\end{equation}
with $T_F=1/2$.

To extract the SDCs of the fragmentation functions, we also need the expressions for the LDMEs of the on-shell $(c\bar{b})$ pair. These LDMEs can be computed from their definitions. The related LDMEs at the LO in $\alpha_s$ are
\begin{eqnarray}
\langle {\cal O}^{(c\bar{b})[^1S^{[8]}_0]}(^1S^{[8]}_0) \rangle &=& N_c^2-1, \\
\langle {\cal O}^{(c\bar{b})[^3S^{[8]}_1]}(^3S^{[8]}_1) \rangle &=& (d-1)(N_c^2-1), \\
\langle {\cal O}^{(c\bar{b})[^1P^{[1]}_1]}(^1P^{[1]}_1) \rangle &=& 2(d-1)N_c \,{\bm q}^2, \\
\langle {\cal O}^{(c\bar{b})[^3P^{[1]}_J]}(^3P^{[1]}_J) \rangle &=& 2(2J+1)N_c \,{\bm q}^2,
\end{eqnarray}
for the $SU_{c}(N_c)$ color group. By utilizing the obtained fragmentation functions and LDMEs of the on-shell $(c\bar{b})$ pair, we derive the SDCs via eq.~(\ref{nrqcd-fact-3}). Subsequently, the fragmentation functions for $P$-wave $B_c$ states are computed using eqs.~(\ref{nrqcd-fact-1}) and (\ref{nrqcd-fact-2}). 

\section{Numerical results and discussion}
\label{numer}

In our numerical calculation, the required input parameters are taken as follows:
\begin{eqnarray}
&& m_c=1.5 ~{\rm GeV}\,,\; m_b=4.9 ~{\rm GeV}\,,\;\vert R'_{2P}(0)\vert^2=0.201~{\rm GeV}^5,
\end{eqnarray}
where the value of $\vert R'_{2P}(0)\vert^2$ is taken from the potential-model (Buchm\"{u}ller-Tye potential) calculation \cite{Eichten:1995ch}. Correspondingly, the mass scale $M$ defined below eq.~(\ref{eq.Pi3}) is explicitly recalled here as $M = m_b + m_c = 6.4~{\rm GeV}$. Here, the $n=2$ states are chosen as a representative case for numerical illustration. It is worth noting that since the SDCs are strictly independent of the principal quantum number $n$ (as discussed in section \ref{calcmethod}), our theoretical framework can be straightforwardly extended to other radial excited $P$-wave states (such as $n=3,4,\dots$) by simply employing their corresponding derivatives of the radial wave functions at the origin.  For the strong coupling constant, we adopt its two-loop running formula, which yields $\alpha_s(2m_c)=0.259$ \cite{Zheng:2019gnb}.

The color-singlet LDMEs are estimated using the HQSS and the relation in eq.~(\ref{eq.csLDMEs}). However, there is no such relation for the color-octet LDMEs. This is because color-octet LDMEs involve transitions of a color-octet quark pair to color-singlet $P$-wave $B_{c}$ states via the emission or absorption of soft gluons. We follow the methodology of ref.~\cite{Yuan:1994hn}: in the limit $\mu_{\Lambda}\gg \mu_{\Lambda 0}$, the perturbative evolution dominates the color-octet LDME contributions. This occurs because the term representing the perturbative evolution in the RGE solution grows logarithmically with the scale $\mu_\Lambda$, eventually overshadowing the initial nonperturbative input. Thus, allowing the initial value in eq.~(\ref{eq.evolsol}) to be neglected. By setting $\mu_{\Lambda}=m_{\rm red}$ and $\alpha_s(\mu_{\Lambda 0})\sim 1$, we obtain:
\begin{eqnarray}
\langle {\cal O}^{B_c(2^1P_1)}(^1S^{[8]}_0) \rangle_{\mu_{\Lambda}=m_{\rm red}} \approx 1.6 \times 10^{-2}\, {\rm GeV}^{3}.
\end{eqnarray}

\begin{figure}[htbp]
\centering
\includegraphics[width=0.48\textwidth]{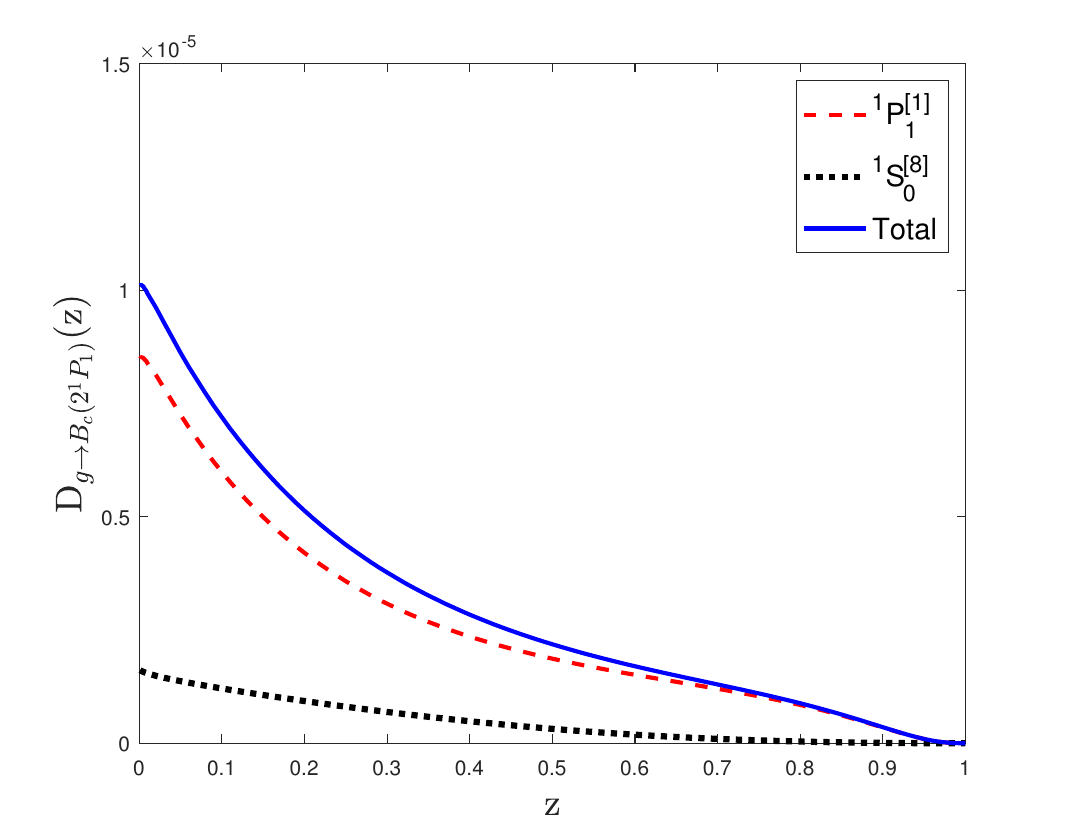}
\includegraphics[width=0.48\textwidth]{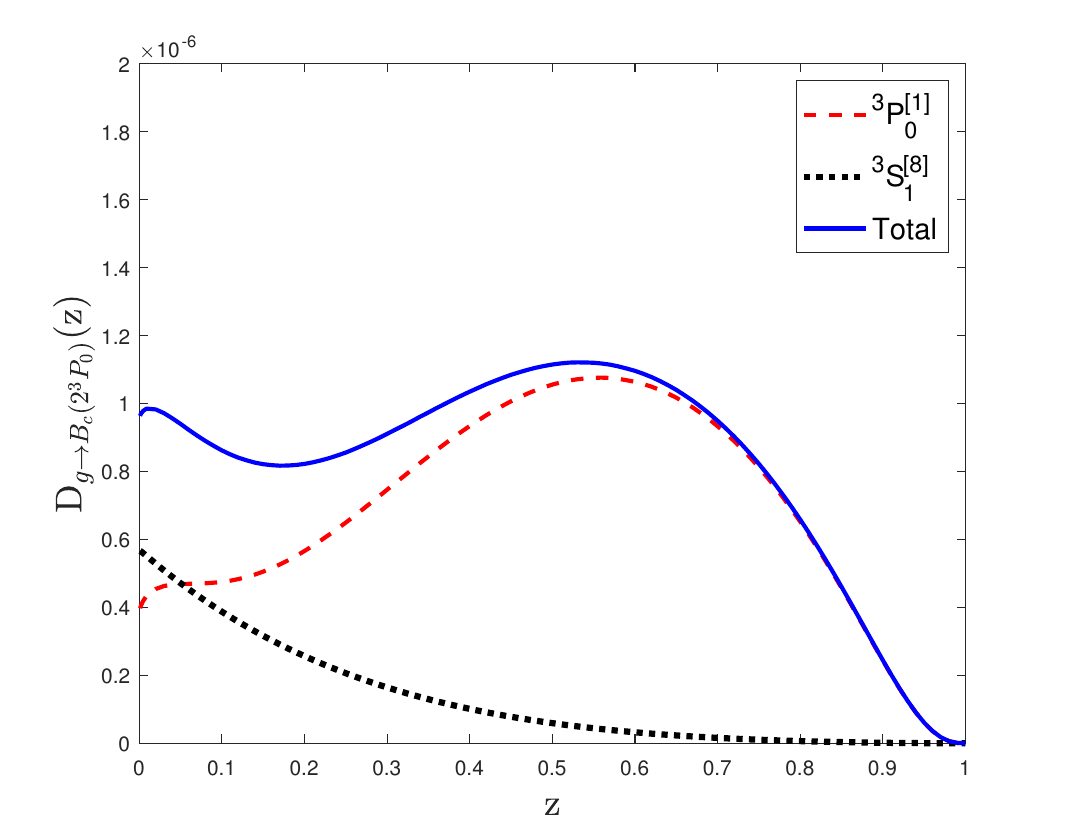}
\includegraphics[width=0.48\textwidth]{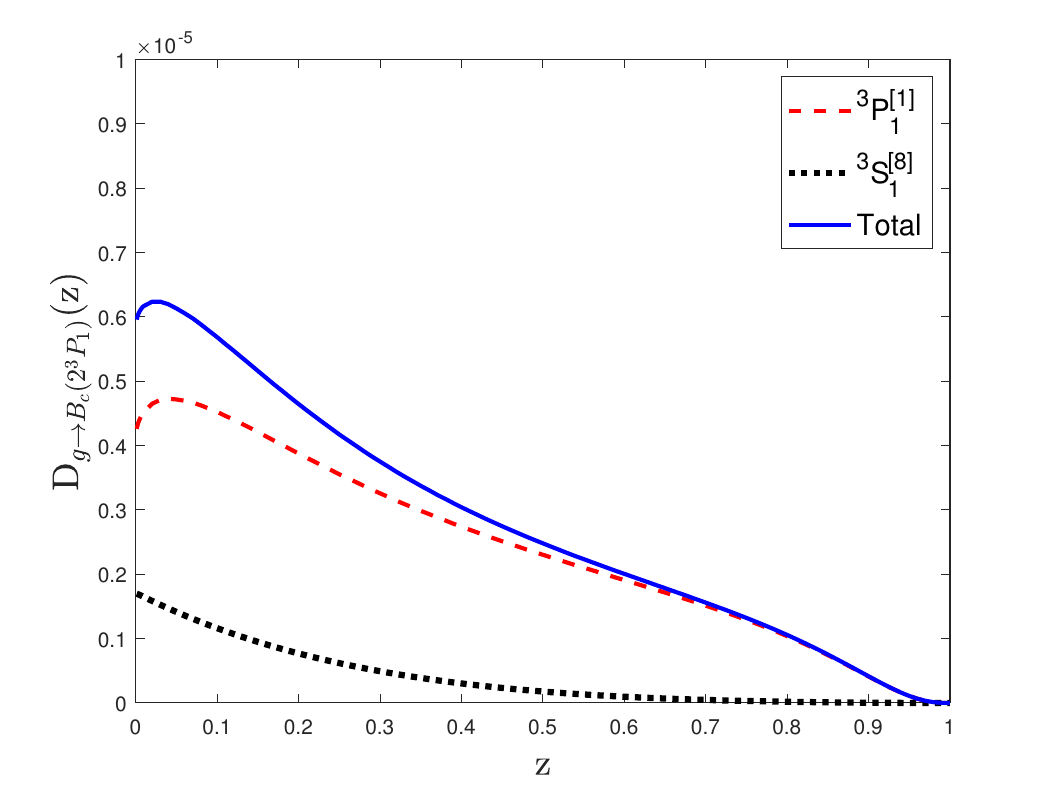}
\includegraphics[width=0.48\textwidth]{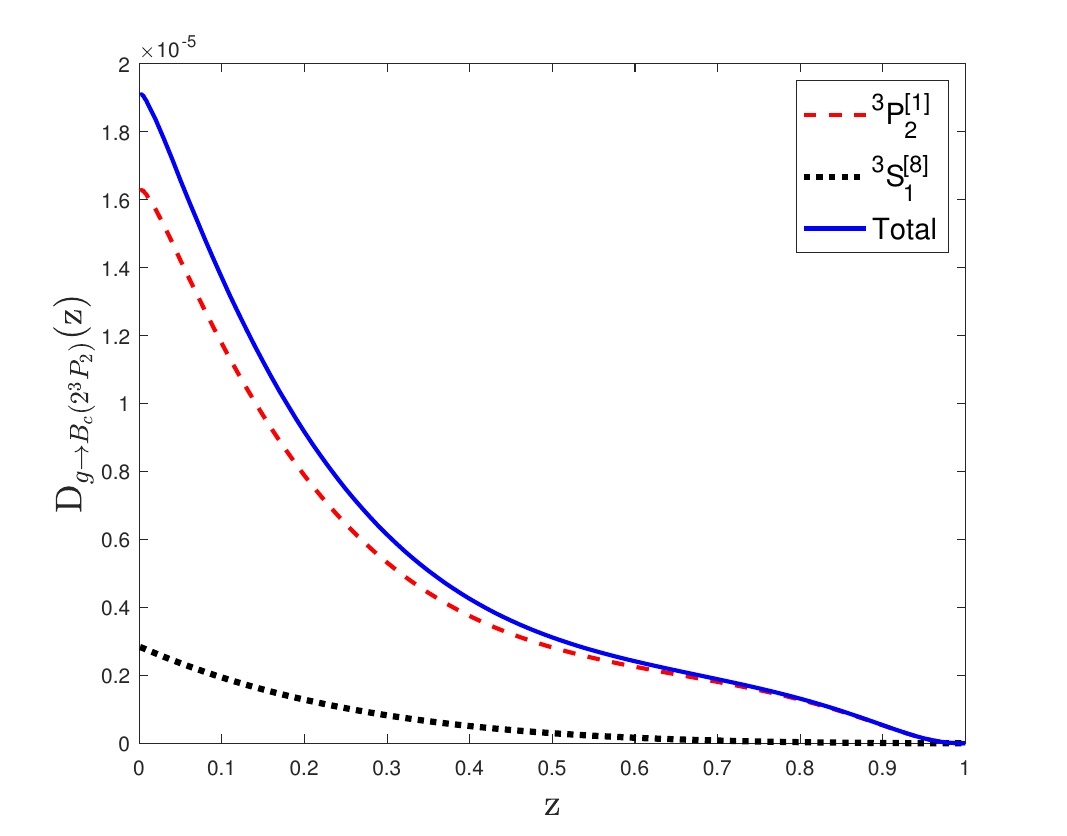}
\caption{Fragmentation functions for a gluon into the $2P$ $B_c$ states as a function of $z$, with $\mu_F=2M$ and $\mu_R=2m_c$.} \label{BcFF}
\end{figure}

In figure \ref{BcFF}, we present the curves of the fragmentation functions $D_{g \to B_c(2^1P_1)}(z,\mu_F)$, $D_{g \to B_c(2^3P_0)}(z,\mu_F)$, $D_{g \to B_c(2^3P_1)}(z,\mu_F)$ and $D_{g \to B_c(2^3P_2)}(z,\mu_F)$, where $\mu_F=2M$ and $\mu_R=2m_c$. The contributions from both color-singlet and color-octet intermediate states are explicitly shown. We can see that, except for the $2\,^3P_0$ state, the color-singlet contribution dominates the fragmentation processes across all relevant $z$ values. For the $2\,^3P_0$ state specifically, the color-singlet contribution remains dominant at large and moderate $z$ values. However, at small $z$ values ($z<0.1$), the color-octet contribution becomes comparable to the color-singlet contribution.

\begin{figure}[htbp]
\centering
\includegraphics[width=0.48\textwidth]{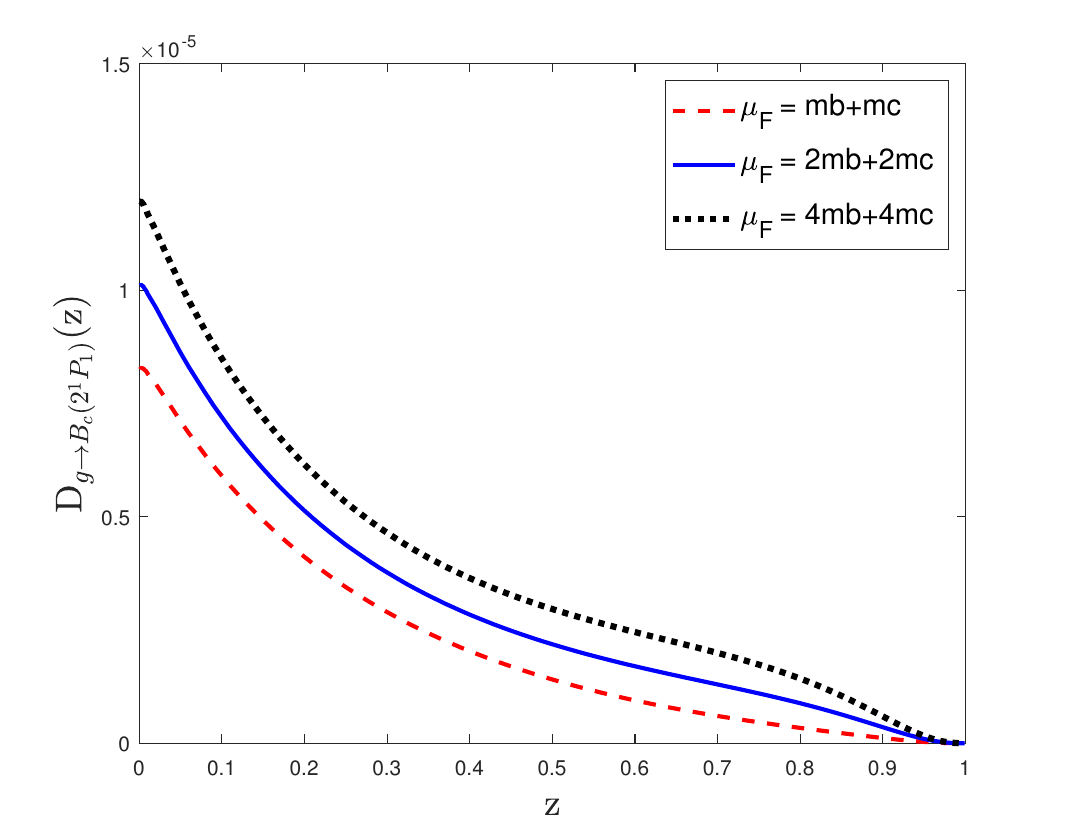}
\includegraphics[width=0.48\textwidth]{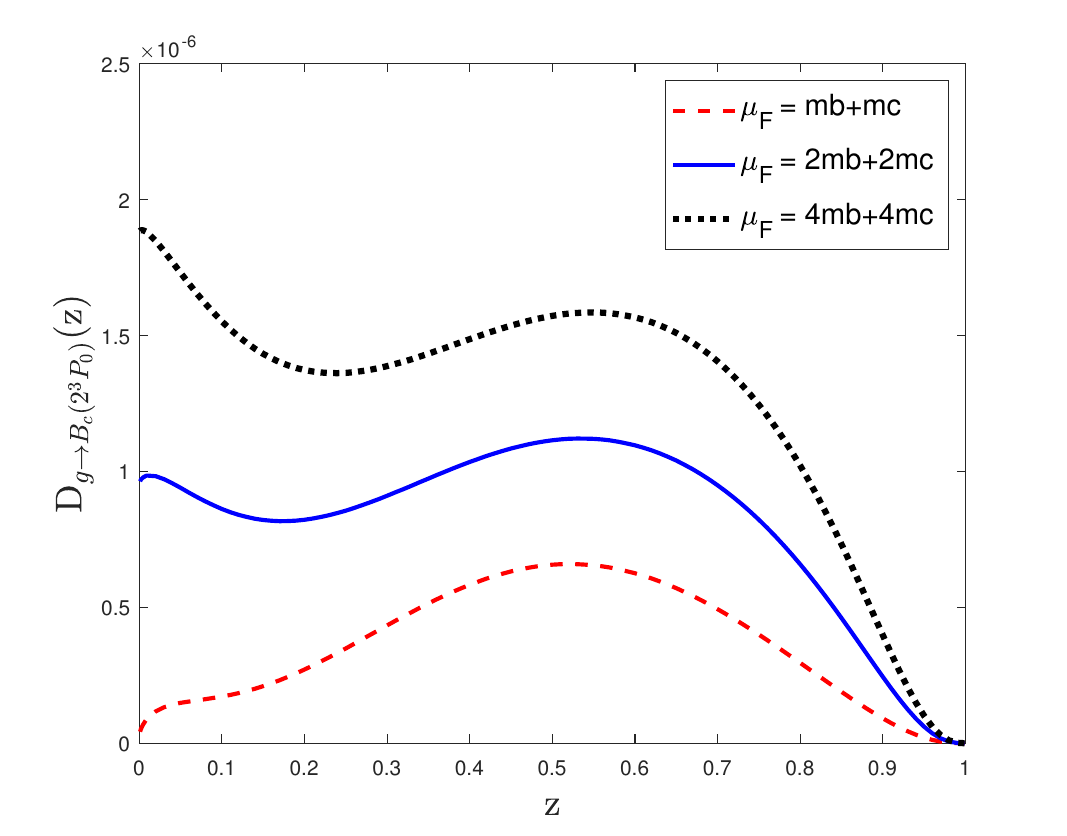}
\includegraphics[width=0.48\textwidth]{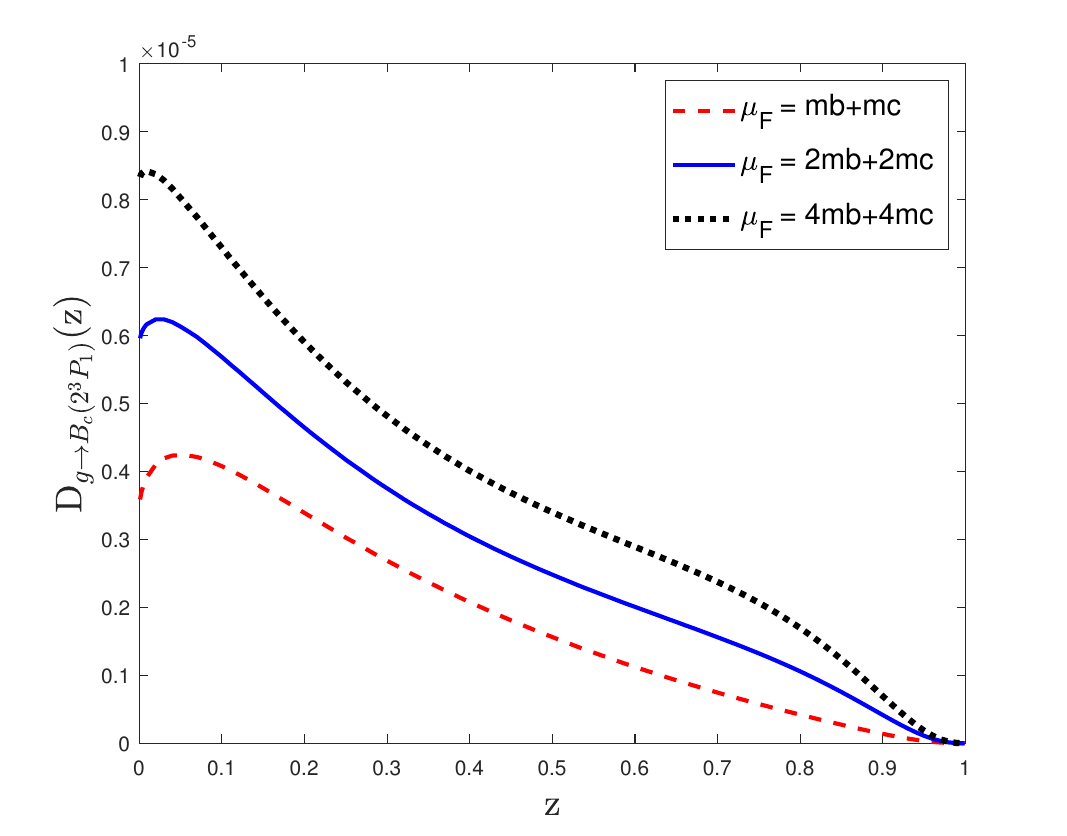}
\includegraphics[width=0.48\textwidth]{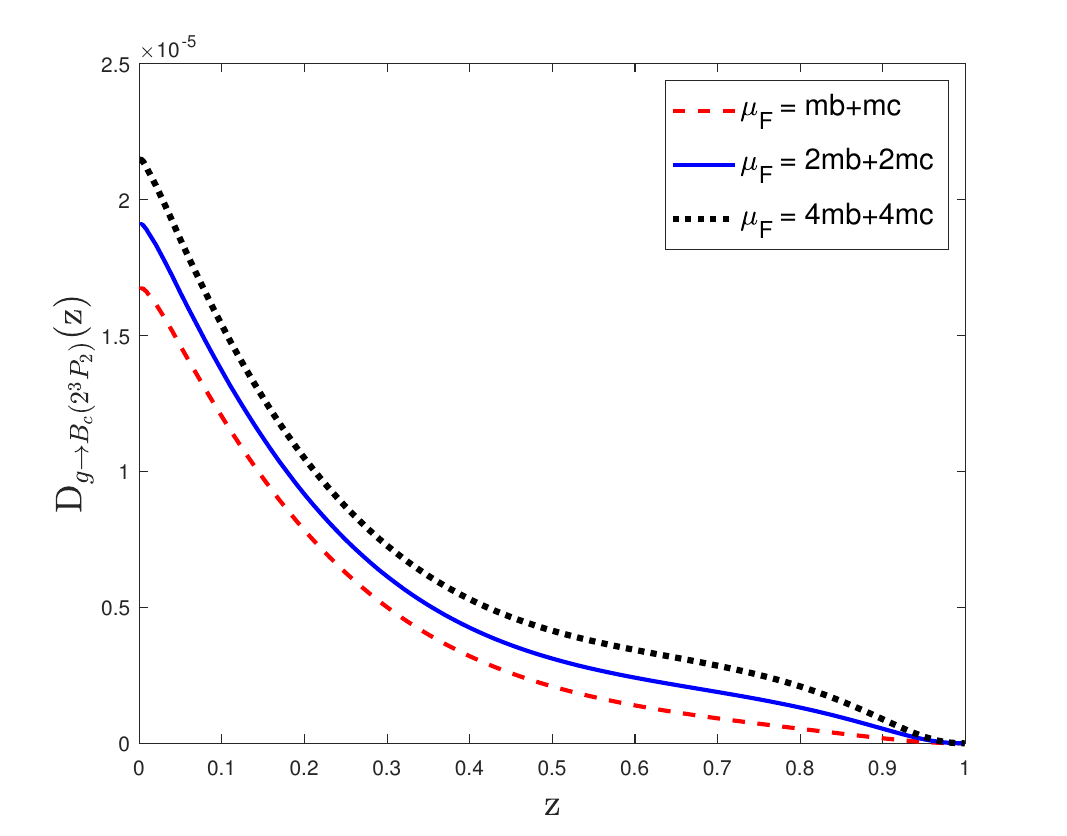}
\caption{Fragmentation functions for $\mu_F=M$, $\mu_F=2M$ and $\mu_F=4M$, where the strong coupling constant is fixed as $\alpha_s(2m_c)=0.259$.} \label{BcFF-muf}
\end{figure}

The fragmentation functions depend on the factorization scale introduced in the renormalization process, as shown in eq.~(\ref{FFRen}). In figure \ref{BcFF-muf}, we plot the curves of the fragmentation functions for $\mu_F=M$, $\mu_F=2M$ and $\mu_F=4M$, respectively. The results indicate that these fragmentation functions are sensitive to the choice of the factorization scale; specifically, the fragmentation functions increase over the entire range $z \in (0,1)$ as the factorization scale rises.

\begin{table}[htb]
\centering
\begin{tabular}{|c c c c| }
\hline
State &  $P(\mu_F=M)$ & $ P(\mu_F=2M)$ & $P(\mu_F=4M)$ \\
\hline
$B_c(2\,^1P_1)$ &  $2.23\times 10^{-6}$ & $3.02\times 10^{-6}$   &  $3.80\times 10^{-6}$ \\
$B_c(2\,^3P_0)$ &  $3.68\times 10^{-7}$ & $8.18\times 10^{-7}$   &  $1.27\times 10^{-6}$ \\
$B_c(2\,^3P_1)$ &  $1.82\times 10^{-6}$ & $2.77\times 10^{-6}$   &  $3.72\times 10^{-6}$ \\
$B_c(2\,^3P_2)$ &  $4.12\times 10^{-6}$ & $5.17\times 10^{-6}$   &  $6.21\times 10^{-6}$\\
\hline
\end{tabular}
\caption{The fragmentation probabilities for $g\to B_c(2\,^{2S+1}P_J)$ with three typical factorization scales, where the strong coupling constant is fixed as $\alpha_s(2m_c)=0.259$.}
\label{tb-fp}
\end{table}

\begin{table}[htb]
\centering
\begin{tabular}{|c c c c| }
\hline
State &  $\langle z \rangle(\mu_F=M)$ & $\langle z \rangle(\mu_F=2M)$ & $\langle z \rangle(\mu_F=4M)$ \\
\hline
$B_c(2\,^1P_1)$ &  0.23 &  0.27  & 0.29 \\
$B_c(2\,^3P_0)$ &  0.49 &  0.44  & 0.42 \\
$B_c(2\,^3P_1)$ &  0.29 &  0.31  & 0.33 \\
$B_c(2\,^3P_2)$ &  0.21 &  0.25  & 0.27 \\
\hline
\end{tabular}
\caption{The average $z$ values for $g\to B_c(2\,^{2S+1}P_J)$ with three typical factorization scales, where the strong coupling constant is fixed as $\alpha_s(2m_c)=0.259$.}
\label{tb-z}
\end{table}

To further characterize the properties of the fragmentation processes, we calculate two key quantities: the fragmentation probability $P$ and the average momentum fraction $\langle z \rangle$. These are mathematically expressed as:
\begin{eqnarray}
P &=& \int_0^1 D(z) dz, \\
\langle z \rangle &=& \frac{\int_0^1 z\,D(z) dz}{\int_0^1 D(z) dz}.
\end{eqnarray}
The numerical results for the fragmentation probabilities and average momentum fractions are summarized in tables \ref{tb-fp} and \ref{tb-z}, where the results for three typical factorization scales are presented. By comparing the data in table \ref{tb-fp} with the results reported in ref.~\cite{Yuan:1994hn}, it can be observed that at the scale $\mu_F={\cal O}(M)$, the probabilities for a gluon to fragment into $P$-wave $B_c$ states are smaller than the corresponding values for the $\bar{b}$-quark fragmentation. However, the gluon fragmentation probabilities are sensitive to the factorization scale and increase as the factorization scale rises. In contrast, the fragmentation probabilities for the $\bar{b}$-quark fragmentation are insensitive to the factorization scale~\cite{Yuan:1994hn}. Furthermore, table \ref{tb-z} shows that the average momentum fraction $\langle z\rangle$ exhibits a weak factorization-scale dependence. To clarify the physical relevance of these behaviors, it should be noted that the fragmentation functions themselves are scheme dependent, unphysical objects defined within the $\overline{\rm MS}$ scheme, and their factorization-scale dependence primarily reflects the evolution trajectories driven by the RGEs. Interestingly, as $\mu_F$ increases, $\langle z\rangle$ grows for the $^1P_1$, $^3P_1$, and $^3P_2$ states, but decreases slightly for the $^3P_0$ state. This distinct behavior is naturally governed by the initial geometric profiles of the fragmentation functions and the $g \to Q\bar{Q}$ ($Q=b,c$) splitting mechanism. Due to the specific scalar projection for the $J=0$ state, the initial distribution of the $^3P_0$ state at $\mu_F = m_b+m_c$ exhibits a relatively broad, flat peak concentrated in the intermediate-to-large $z$ region while being strictly suppressed in the lower-$z$ domain, yielding a higher initial $\langle z\rangle$. When evolved to higher scales via the RGEs, the scale dependence is predominantly driven by the convolution between the splitting function $P_{Qg}$ and the LO heavy-quark fragmentation functions $D_{Q\rightarrow(c\bar{b})[n]}^{\rm LO}$ (see eq.~(\ref{FFRen})). This process structurally generates a strong feeding effect that rapidly enhances the lower-$z$ domain, which significantly shifts the physical weight balance of the $^3P_0$ state leftward and results in a decreasing $\langle z\rangle$. This phenomenon underscores the rich spin-dependent QCD dynamics embedded in the $P$-wave fragmentation processes.

For the convenience of future phenomenological applications, we present analytic fits to the calculated SDCs. As emphasized before, these fitted SDCs carry no dependence on the principal quantum number $n$, meaning that the parameterized functions and the extracted parameters listed below can be directly applied to predict the fragmentation functions of higher $P$-wave $B_c$ states as well. The analytical expression can be written as follows:
\begin{eqnarray}
d_{g\to (c\bar{b})[^{2S+1}L^{[1(8)]}_J]}(z,\mu_{F}) &=& \frac{\alpha_s}{2\pi}{\rm ln}\frac{\mu_{F}^2}{4M^2}\int_z^1 \Bigg[\sum_{Q=\bar{b},c}P_{Qg}(y) d_{Q\to (c\bar{b})[^{2S+1}L^{[1(8)]}_J]}^{\rm LO}(z/y)\Bigg]\frac{dy}{y} \nonumber\\
&& +\frac{\alpha_s^3}{M^{(2L+3)}} f_{[^{2S+1}L^{[1(8)]}_J]}(z),
\label{eqDzfit}
\end{eqnarray}
where $L=0,1$ or $S$, $P$ waves. The expressions of the LO SDCs $d_{Q\to (c\bar{b})[^{2S+1}L^{[1(8)]}_J]}^{\rm LO}(z)$ are presented in appendix \ref{Ap.SDCs}. We parametrize the functions $f_{[^{2S+1}L^{[1(8)]}_J]}(z)$ as
\begin{eqnarray}
f_{[^{2S+1}L^{[1(8)]}_J]}(z)=N\,z^{\alpha}(1-z)^{\beta} \left( 1+\sum_{i=1}^{6}a_i\, z^i \right).
\label{eqDzfit2}
\end{eqnarray}
The fitted parameters for different SDCs are given in table \ref{tb-fit}. Using the factorization formulas in eqs.~(\ref{nrqcd-fact-1}) and (\ref{nrqcd-fact-2}), one can easily reproduce the fitted fragmentation functions for the $P$-wave $B_c$ states.


\begin{table}[htb]
\centering
\begin{tabular}{|c c c c c c c c c c| }
\hline
State         &  $N$    & $\alpha$ & $\beta$ & $a_1$    & $a_2$ & $a_3$   & $a_4$  & $a_5$  & $a_6$  \\
\hline
$^1P_1^{[1]}$ & 6.705  &  0.01418  & 2.172   & -2.293   & 7.423 & -18.93  & 47.56  & -65.72 & 41.25   \\
$^3P_0^{[1]}$ & 1.297  &  0.06646  & 2.269   & 0.7862   & 4.369 & 34.07   & 140.1  & -381.9 & 336.1   \\
$^3P_1^{[1]}$ & 3.753  &  0.03236  & 2.138   & 2.702    & -15.01& 51.22   & -66.48 & 26.77  & 18.66   \\
$^3P_2^{[1]}$ & 7.580  &  0.01234  & 2.114   & -1.318   & -2.371& 14.21   & -17.72 & 4.866  & 8.175   \\
$^1S_0^{[8]}$ & 1.420  &  -0.01096 & 3.101   & 0.3302   & 7.717 & -33.55  & 86.58  & -108.5 & 56.07   \\
$^3S_1^{[8]}$ & 1.619  &  0.001004 & 1.675   & -2.113   & 2.244 & -2.408  & 2.907  & -2.043 & 0.4914  \\
\hline
\end{tabular}
\caption{The fitted parameters for the functions $f_{[^{2S+1}L^{[1(8)]}_J]}(z)$ defined in eq.~(\ref{eqDzfit2}).}
\label{tb-fit}
\end{table}

\section{Summary}
\label{sum}

In this study, we have derived the fragmentation functions for a gluon into $P$-wave $B_c$ mesons, which start at order $\alpha_s^3$. Our analysis includes contributions from both color-singlet and color-octet intermediate states.

The phase-space integrals exhibit UV divergences, which are addressed using dimensional regularization. Direct evaluation of these integrals in $d$-dimensional spacetime proves challenging; therefore, we employ a subtraction method to handle the UV divergent integrals, allowing us to accurately determine both the divergent and finite contributions. The UV divergences are managed through the renormalization of the operator definitions of the fragmentation functions within the $\overline{\rm MS}$ scheme.

The gluon fragmentation functions into $P$-wave $B_c$ mesons, defined in the $\overline{\rm MS}$ factorization scheme, are presented in both graphical form and as analytic fits. We observe a pronounced dependence of the fragmentation functions on the factorization scale. In particular, the fragmentation functions increase over the entire region $0<z<1$ as the factorization scale becomes larger, indicating that gluon fragmentation plays an increasingly important role in the $P$-wave $B_c$ production at very high energies. We also present the fragmentation probabilities and the average momentum fractions $\langle z \rangle$ at three representative factorization scales. The results indicate that, although the fragmentation probabilities depend strongly on the factorization scale, the average values of $z$ show only a very weak scale dependence. Interestingly, the fragmentation probabilities remain positive even at the scale $\mu_F = m_b + m_c$, which lies below the kinematic threshold for a gluon fragmentating into a $P$-wave $B_c$ state. Our results can serve as inputs for precision phenomenological studies of $P$-wave $B_c$ meson production at high-energy collider experiments.

\acknowledgments
This work was supported in part by the Natural Science Foundation of China under Grants No. 12575080 and No. 12547101, and by the Chongqing Natural Science Foundation under Grant No. CSTB2025NSCQ-GPX0745. \\


\appendix

\section{LO SDCs for the fragmentation functions of a $\bar{b}(c)$ quark into $P$-wave $B_c$ mesons}
\label{Ap.SDCs}

In this paper, we present analytic fits for the SDCs related to the fragmentation functions of a gluon into $P$-wave $B_c$ mesons, as outlined in eqs.~(\ref{eqDzfit}) and (\ref{eqDzfit2}). These fitted SDCs require the SDCs for the fragmentation functions of a $\bar{b}(c)$ quark into $P$-wave $B_c$ mesons as inputs. The leading-order (LO) SDCs for the fragmentation functions of a $\bar{b}(c)$ quark into $P$-wave $B_c$ mesons have been calculated in refs.~\cite{Chen:1993ii,Yuan:1994hn}. For convenience, we provide these SDCs here. The SDCs for the $\bar{b}$-quark fragmentation in $d=4$ dimensions are
\begin{eqnarray}
d^{\rm LO}_{\bar{b}\to (c\bar{b})[^1P_1^{[1]}]}(z)=&& \frac{\alpha_s^2 z(1-z)^2}{729 M^5 r_b^2 r_c^4(1-r_b z)^8} [6-6(5-8r_c+4r_c^2)z+(69-210r_c+250r_c^2 \nonumber \\
 &&-96r_c^3 +32r_c^4)z^2-8 r_b(12-37r_c+48r_c^2-12r_c^3-4r_c^4)z^3+2r_b^2(42 \nonumber \\
 && -114r_c+161r_c^2+16r_c^4)z^4-6r_b^3(7-15r_c+28r_c^2+4r_c^3)z^5+r_b^4(9 \nonumber \\
 && -14r_c+46r_c^2)z^6],\\
 d^{\rm LO}_{\bar{b}\to (c\bar{b})[^3P_0^{[1]}]}(z)=&& \frac{\alpha_s^2 z(1-z)^2}{729 M^5 r_b^2 r_c^4(1-r_b z)^8} [6(1-4r_c)^2-6(1-4r_c)(5-16r_c+20r_c^2)z \nonumber \\
 &&+(63-362r_c+1058r_c^2-1456r_c^3 +832r_c^4)z^2-8 r_b(9-22r_c+118r_c^2 \nonumber \\
 && -184r_c^3+100r_c^4)z^3+2r_b^2(24+42r_c+369r_c^2-776r_c^3+416r_c^4)z^4 \nonumber \\
 && -2r_b^3(9+59r_c+232r_c^2-516r_c^3+240r_c^4)z^5+r_b^4(3+34r_c+134r_c^2 \nonumber \\
 &&-240r_c^3+96r_c^4)z^6],\\
 d^{\rm LO}_{\bar{b}\to (c\bar{b})[^3P_1^{[1]}]}(z)=&& \frac{2\alpha_s^2 z(1-z)^2}{729 M^5 r_b^2 r_c^4(1-r_b z)^8} [6-6(5-4r_c)z+(63-98r_c+64r_c^2 \nonumber \\
 && +16r_c^4)z^2-8 r_b(9-11r_c+13r_c^2-2r_c^3-2r_c^4)z^3+2r_b^2(24-18r_c \nonumber \\
 && +47r_c^2-16r_c^3+8r_c^4)z^4 -2r_b^3(9-r_c+24r_c^2-8r_c^3)z^5+r_b^4(3+2r_c\nonumber \\
 && +12r_c^2)z^6], \\
 d^{\rm LO}_{\bar{b}\to (c\bar{b})[^3P_2^{[1]}]}(z)=&& \frac{4\alpha_s^2 z(1-z)^2}{3645 M^5 r_c^4(1-r_b z)^8} [12-12(5-2r_c)z+(135-76r_c+92r_c^2)z^2 \nonumber \\
 &&-4(45-31r_c+54r_c^2-10r_c^3)z^3+2(75-78r_c+123r_c^2-16r_c^3 \nonumber \\
 && +46r_c^4)z^4-4r_b(18-13r_c+40r_c^2+9r_c^3+6r_c^4)z^5+r_b^2(15-10r_c \nonumber \\
 && +55r_c^2-12r_c^3+12r_c^4)z^6],\\
 d^{\rm LO}_{\bar{b}\to (c\bar{b})[^1S_0^{[8]}]}(z)=&& \frac{\alpha_s^2 z(1-z)^2}{648 M^3 r_c^2(1-r_b z)^6} [6-18(1-2r_c)z+(21-74r_c+68r_c^2)z^2 \nonumber \\
 &&-2r_b(6-19r_c+18r_c^2)z^3+3r_b^2(1-2r_c+2r_c^2)z^4 ],\\
 d^{\rm LO}_{\bar{b}\to (c\bar{b})[^3S_1^{[8]}]}(z)=&& \frac{\alpha_s^2 z(1-z)^2}{648 M^3 r_c^2(1-r_b z)^6} [2-2(3-2r_c)z+3(3-2r_c+4r_c^2)z^2 \nonumber \\
 &&-2r_b(4-r_c+2r_c^2)z^3+r_b^2(3-2r_c+2r_c^2)z^4 ].
\label{eqa.Dzfit}
\end{eqnarray}
The SDCs for $c \to (c\bar{b})[^{2S+1}L_J^{[1(8)]}]$ can be obtained from the corresponding SDCs for $\bar{b} \to (c\bar{b})[^{2S+1}L_J^{[1(8)]}]$ through the exchange $r_b \leftrightarrow r_c$.

\providecommand{\href}[2]{#2}\begingroup\raggedright

\end{document}